# Continuous Focus Groups: A Longitudinal Method for Clinical HRI in Autism Care*

Ghiglino D., Foglino C., Wykowska A.

*Abstract*— Qualitative methods are important to use alongside quantitative methods to improve Human–Robot Interaction (HRI), yet they are often applied in static or one-off formats that cannot capture how stakeholder perspectives evolve over time. This limitation is especially evident in clinical contexts, where families and patients face heavy burdens and cannot easily participate in repeated research encounters. To address this gap, we introduce continuous focus groups, a longitudinal and co-agential method designed to sustain dialogue with assistive care professionals working with children with autism spectrum disorder (ASD). Three focus groups were organized across successive phases of a robot-assisted therapeutic protocol, enabling participants to revisit and refine earlier views as the intervention progressed. Results show that continuity fostered trust, supported the integration of tacit clinical expertise into design decisions, and functioned as an ethical safeguard by allowing participants to renegotiate involvement and surface new concerns. By bridging the therapeutic iteration of families, children, and clinicians with the research–design iteration of researchers, developers, continuous focus groups provide a methodological contribution that is both feasible in practice and rigorous in design. Beyond autism care, this approach offers a transferable framework for advancing qualitative research in HRI, particularly in sensitive domains where direct user participation is limited and continuity is essential.

## I. INTRODUCTION

The field of Human–Robot Interaction (HRI) is characterized by its interdisciplinarity and its persistent methodological challenges. While it has established itself as a distinct research area over the last two decades, debates remain about how to ensure rigor, comparability, and relevance in studies that necessarily involve complex social and clinical contexts (Veling & McGinn, 2021). Calls for methodological pluralism have been made repeatedly, stressing that the field should not be restricted to quantitative paradigms derived from psychology and engineering, but should also integrate qualitative and interpretive approaches that capture the lived experiences of participants (Lim, 2025). These approaches are particularly valuable in clinical environments, where rigidly controlled studies may fail to capture the emergent and context-dependent dynamics of human care practices. It is within this space that the present study is situated. The work forms part of the RObot traiNing INdependence "RONIN" project, investigating how humanoid robots can support the development of autonomy and socio-cognitive skills in children with autism spectrum disorder (ASD). In this context, the primary interlocutors of our research were clinicians working in a rehabilitation center, who bring specialized expertise in translating therapeutic goals into practice and interpreting children's communicative behaviors. Qualitative research in HRI has expanded significantly, with focus groups occupying an important role among the available methodologies. They have been employed to capture users' and stakeholders' perceptions of robots in varied contexts, including education, health care, eldercare, and autism-related interventions. For example, Krupp et al. (2017) demonstrated how focus groups can reveal subtle concerns about privacy and ethics in the use of telepresence robots in healthcare. In another study, Smakman, Vogt, & Konijn (2021) used multi-stakeholder focus groups to highlight moral divergences regarding robots in educational settings, showing how the same technology may be perceived as empowering by some actors and threatening by others. Also Camilleri, Dogramadzi, & Caleb-Solly (2022) reported that caregivers' tacit knowledge, often invisible in design processes, can be elicited through group discussion, thus reshaping the design priorities of socially assistive robots. Similarly, Bradwell et al. (2021) illustrated how focus groups with older adults and care staff expose a mismatch between developers' assumptions and end-users' lived realities. These studies underscore the value of focus groups in uncovering perspectives that might otherwise remain overlooked. Yet they also share an important limitation: they are often designed as one-off events, providing snapshots of stakeholder views but not sustained engagement or iterative dialogue.

Beyond HRI, research in clinical and caregiving contexts reinforces the necessity of sustained, multiprofessional communication. Studies on autism care demonstrate that families and caregivers frequently face fragmented systems and are forced to assume an advocacy role to bridge gaps in support (Bonfim et al., 2023; Mazurek et al., 2021). Guler et al. (2023) showed, for instance, that logistical and structural barriers hinder consistent caregiver engagement in autism interventions. These findings point to the risks of superficial consultation: without continuity and integration, stakeholders' insights remain isolated, and systemic gaps persist. At the same time, it is important to recognize that both patients and caregivers often live under significant burdens. Families caring for autistic children, for instance, may already feel overwhelmed by the demands of daily support, making it neither realistic nor ethical to ask them to sustain repeated involvement in research. Moreover, children with ASD —

*Research supported by ERC Proof of Concept program.
G. D. is with the Social Cognition in Human-Robot Interaction, Italian Institute of Technology, Genova, 16163 Italy. (e-mail: davide.ghiglino@iit.it).

F. C. is with the Social Cognition in Human-Robot Interaction, Italian Institute of Technology, Genova, 16163 Italy. (e-mail: caterina.foglino@iit.it).

W. A. is with the Social Cognition in Human-Robot Interaction, Italian Institute of Technology, Genova, 16163 Italy (corresponding author; phone: +39 010 2897 242; e-mail: agnieszka.wykowska@iit.it).

especially those with limited or atypical verbal abilities — cannot reliably participate in group discussions that depend on dialogue. For these reasons, the methodological choice was made to focus continuous focus groups on clinicians. Therapists are embedded in care practices, possess both tacit and formalized knowledge, and unlike families, they are better positioned to engage consistently in longitudinal research encounters.

At the methodological level, the debate around rigor in qualitative research also intersects with this challenge. Lim (2024), for example, emphasizes that rigor in qualitative inquiry cannot be reduced to replication or statistical generalizability; instead, it must be pursued through credibility, reflexivity, transferability, and continuous consent. Taquette and Souza (2022) extend this perspective by drawing attention to the ethical dilemmas that arise in qualitative settings, particularly when working with vulnerable groups. They argue that the researcher's position and the ongoing negotiation of autonomy and confidentiality are not peripheral but central to methodological soundness. This ethical dimension is of particular relevance in clinical HRI, where the inclusion of therapists, caregivers, and patients entails heightened responsibilities.

Taken together, these bodies of work establish both the promise and the shortcomings of qualitative methods in HRI. Focus groups have proven effective at generating insights into user needs, ethical perceptions, and design priorities, yet their application has often been one-off and disconnected from longitudinal collaboration. Clinical and caregiving studies demonstrate the systemic necessity of continuous multiprofessional dialogue but also reveal the difficulty of achieving it in practice. Furthermore, the everyday strain experienced by families and patients underscores the importance of identifying continuity among interlocutors who can sustain engagement without adding to the burdens of care. Against this backdrop, there is a clear need for methodological frameworks that combine continuity, interdisciplinarity, and co-agency (i.e., one that emphasizes shared agency between clinicians and researchers in shaping the process), particularly in sensitive contexts such as autism care, where the stakes are both clinical and social.

This paper addresses this gap by introducing and evaluating continuous focus groups as a methodological contribution to HRI research. By framing focus groups not as isolated consultations but as iterative, evolving dialogues, we demonstrate how therapists' clinical expertise can be integrated into the design and reflection on robot-assisted interventions. In doing so, we argue that continuous focus groups constitute not only a methodological refinement but also an ethical and epistemological stance: one that treats clinicians as co-designers in the research process, acknowledges the emergent character of care practices, and advances the field of HRI toward more grounded and context-sensitive forms of inquiry.

Beyond continuity and interdisciplinarity, another critical dimension concerns communication across domains. Clinical expertise and technological development are often articulated in distinct professional languages, with different assumptions, constraints, and priorities. For collaboration to be effective, there is a need for intermediary figures capable of "speaking both languages"—that is, of translating clinical requirements into technical specifications and technological constraints back into therapeutic terms. The presence of such mediators is essential to ensure that iterative processes between clinical and technical worlds remain productive and grounded in the realities of both care and design.

## II. MATERIALS AND METHODS

### A. Context

The study was carried out in collaboration with a local rehabilitation center that has an established history of working with our research group on robot-assisted therapy (RAT). Over the years, this collaboration has enabled the introduction of humanoid robots directly into clinical routines, fostering familiarity and trust among staff and ensuring that research activities could be smoothly integrated into therapeutic practice. This long-standing partnership created the conditions for organizing not only interventions but also prolonged and continuous sessions, including structured opportunities for feedback such as focus groups.

For the present study, three focus groups were conducted at distinct phases that paralleled the crossover design commonly adopted in RAT research (see Figure 1).

Each session followed a block of clinical activities in which a humanoid robot was directly involved with children with autism spectrum disorder (ASD) under the supervision of care professionals. In both phases of the intervention, children interacted repeatedly with the robot over a period of approximately two months, engaging in structured training tasks based on building sequences of actions with cubes. Each cube depicted an element of a routine or scenario, and together they formed short stories designed to elicit reasoning about everyday situations. Importantly, the sequences were built collaboratively: the robot and the child alternated turns, exchanging cubes through handovers that incorporated social cues such as gaze and mutual attention (see Figure 2). In the first phase, the sequences represented self-care routines (e.g., brushing teeth: selecting a toothbrush, adding toothpaste, brushing). In the second phase, the focus shifted to social interaction scenarios, such as helping another person in distress.

This adjustment followed therapist feedback during the second focus group, where it emerged that many children mastered autonomy-related routines after only a few sessions and began to lose interest. For children with higher cognitive skills, additional challenge was created by introducing distractor cubes—either irrelevant actions or conceptually wrong ones (e.g., raging at a person in distress)—and by having the robot itself sometimes make intentional mistakes for the child to detect. To prevent habituation and carry-over effects, the specific actions and sequences were varied in every session. All tasks followed a two-session structure: in the first, the robot initiated the sequence; in the second, the child took the lead. At the end of each session, children were prompted to verbally summarize the completed routine.

Prior to the first focus group, a demonstration session was also organized for therapists as a familiarization phase. Here, children built neutral and playful stories (e.g., a dog playing with a cat in a park) using three cubes whose order was not predetermined. The goal was not correctness but to allow children to get accustomed to the robot's behavior and the general structure of the task. As the detailed design of the training is not the main focus of this paper, we report it here only briefly to provide context for the focus group discussions.

The focus groups took place physically inside the rehabilitation center, immediately after the intervention phases, and were designed to capture clinicians' evolving reflections while their experience with the robot was fresh and situated in practice. This design allowed us to link qualitative insights directly to the progression of the intervention.

The participants were assistive care professionals—including psychologists, speech therapists, psychomotricists, and educators—who were already part of the therapeutic teams running the sessions with children. Their role in the focus groups was therefore informed not only by professional expertise but also by their direct, recent experience of working with the robot in therapy. Families were indirectly engaged in the clinical activities, as clinicians' feedback reflected a natural iteration of perspectives that emerged in daily exchanges among therapists, children, and caregivers. This triangular dynamic is a defining feature of rehabilitation practice and provided additional depth to the reflections shared in the groups.

Although in this article we report specifically on the three focus groups connected to the present intervention, it is important to note that group-based feedback collection has been a recurring practice in our collaborations with the center. What distinguishes the current work is the recognition of these sessions as a structured methodological approach, rather than as ancillary or anecdotal feedback. By systematically analyzing how focus groups conducted at multiple time points can generate longitudinal and clinically informed insights, we highlight their methodological value for HRI research. This is particularly relevant in a field where qualitative data are often underreported, despite their pivotal role in contextualizing and complementing quantitative measures.

### B. Procedure

Three focus groups were conducted at distinct phases of the rehabilitation program, each aligned with a different stage of the robot-assisted intervention. All sessions were held in a dedicated meeting room inside the rehabilitation center and scheduled immediately after the clinical activities, in order to maintain continuity between therapeutic practice and reflective discussion.

The first focus group took place in July 2024, following a demonstration session during which the humanoid robot iCub was introduced to both therapists and children. This preliminary step ensured that all professionals had the opportunity to see the robot in action before the formal therapeutic training was initiated. On that occasion, twenty-four therapists attended the group. The second focus group was organized in January 2025, after the first intervention phase in which an initial group of children with autism spectrum disorder underwent structured robot-assisted training. This session brought together sixteen therapists, who were able to reflect on their direct experience of the robot within therapeutic routines and to comment on the children's reactions and families' perspectives. The third focus group was held in May 2025, following the second intervention phase. In this cycle, the children who had not participated in the earlier round of training were engaged in an enhanced version of the protocol, which had been refined on the basis of the feedback collected during the previous group (Figure 1 serves as a reference for the crossover design adopted in the current study). For this final session, twenty therapists took part.

In sum, 24 professionals took part in the groups, though the composition of the groups varied between timepoints according to their availability. All participants were female, with an age range between twenty-four and fifty-one years. All had several years of experience in neurodevelopmental disorders and were directly involved in the robot-assisted activities with children. Their prior familiarity with the iCub robot and with the research team, built through previous collaborations, allowed the discussions to focus immediately on substantive clinical and technical feedback rather than introductory or "warm-up" activities. Each session lasted between sixty and ninety minutes and was facilitated by two researchers with complementary expertise in clinical psychology and human–robot interaction. Participation was always voluntary, and while attendance was noted for reporting purposes, no record was kept of individual contributions. At the start of every session, the facilitators welcomed participants, invited them to pause before beginning, and reminded them that the group was intended as a safe and non-judgmental space. Both positive and critical feedback were explicitly encouraged, and confidentiality was guaranteed. This setting was designed to reduce pressure, acknowledge the considerable clinical workload of participants, and foster an atmosphere where free expression was possible.

The groups followed a semi-structured format. Discussions opened with a round of general impressions about the robot-assisted protocol. As certain themes recurred, the facilitators highlighted them, encouraged agreement or disagreement, and invited alternative perspectives, thus ensuring that diverse viewpoints could emerge. In the subsequent phase of dialogue, participants were asked to identify what they perceived as the main strengths and weaknesses of the protocol. Particular attention was directed toward two dimensions: the procedural aspects, including the structure of stimuli, sequence of robot prompts, and therapeutic relevance; and the technical aspects, such as the fluidity of movements, timing of responses, stability of behaviors, and appropriateness of voice and speech. The facilitators actively supported the discussion by summarizing points raised, proposing possible adaptations or solutions, and checking whether these aligned with the collective views of the group.

This balance between open dialogue and targeted prompts enabled participants to articulate feedback in their own words while ensuring that the conversation remained relevant to the aims of the study. Each session concluded with a short debriefing, in which the facilitators recapped the main themes that had emerged, followed by a coffee break offered as a gesture of appreciation and to reinforce the collaborative and informal atmosphere of the encounter.

### C. Facilitation

The focus groups were facilitated by a clinical psychologist who had a complementary background in HRI and psychotherapy, with two additional co-moderators trained in psychotherapy and rehabilitation. This background proved valuable in establishing rapport with participants and in maintaining sensitivity to the professional and emotional context in which feedback was elicited. Rather than following a rigid script, moderators adopted a dual role: guiding the discussion toward topics relevant for the research aims while also attending to the relational climate of the group. Active listening techniques, such as paraphrasing participants' remarks and explicitly acknowledging divergent views, were used to validate contributions and sustain engagement.

An additional role of the facilitators was to mediate between disciplinary vocabularies. Therapists often framed their observations in clinical terms—such as the child's engagement, attention span, or communicative initiative—while the research team required feedback that could be translated into technical specifications, for example timing of robot responses or movement stability. The facilitators worked to bridge these perspectives, ensuring that clinicians' suggestions could be reformulated into actionable design input without losing their original meaning.

Finally, attention was paid to group dynamics. Moderators monitored participation levels to prevent the discussion from being dominated by a few voices, while leaving ample room for spontaneous exchange. By preserving a balance between structure and openness, facilitation contributed not only to the quality of the data collected but also to the progressive increase in trust and expressiveness observed across sessions. In this sense, the facilitators' role can be seen as embodying the broader function of interdisciplinary mediation. Their capacity to reformulate clinicians' observations into terms intelligible for developers, while preserving the original therapeutic meaning, illustrates the importance of professional figures who operate at the intersection of clinical practice and technological design. This role extends beyond facilitation of dialogue, pointing to a structural need in HRI research for individuals trained to bridge domains systematically.

*D. Data Collection and Analysis*

All focus group discussions were audio-recorded and transcribed in full, and detailed notes were taken by the facilitators to capture non-verbal dynamics and contextual elements. In addition, observations from the pilot sessions and subsequent fieldwork with children were integrated into the analysis, providing a richer backdrop against which therapists' feedback could be interpreted. The analytic process followed a thematic synthesis approach, moving iteratively between the transcripts and the field notes to identify recurrent themes as well as points of divergence. Particular attention was devoted to two dimensions: on the one hand, the methodological insights that emerged regarding the structure, dynamics, and utility of iterative focus groups for HRI research; on the other, the theoretical implications concerning how diverse forms of expertise—clinical, technical, and research—can be integrated in the development of robotic interventions. This dual lens made it possible to capture both the practical value of the focus groups and their broader significance for advancing methods and theory in HRI. A schematic overview of the iterative processes underlying the study is presented in Figure 3, which illustrates how continuous focus groups served as a bridge between the clinical–therapeutic iteration and the research–design iteration.

III. RESULTS

*A. Focus Group 1: Exploratory Phase*

The first focus group, held after the demonstration sessions, was largely exploratory. Therapists discussed the initial structure of the protocol and emphasized the need to adapt task complexity to the heterogeneous profiles of the children. For some participants, the proposed sequences were too demanding and risked producing frustration; for others, the tasks appeared overly simple and repetitive. The group suggested scaffolding strategies such as reducing distractors for children with lower attentional resources, while increasing variability for those with higher competence. In addition to task complexity, the therapists highlighted the importance of addressing domains of autonomy and social interaction, proposing scenarios such as self-care routines and basic social exchanges. These discussions made clear that the protocol should balance therapeutic aims with children's everyday contexts.

*B. Focus Group 2: Adaptive Phase*

The second focus group took place after the first round of revisions had been implemented and tested in practice. The discussion focused on satisfaction with the modifications, while also opening space for further refinement. Therapists reported progress in tailoring activities, but they continued to stress the centrality of robot behavior. Movements were perceived as too slow or mechanical, sometimes leading to disengagement. Suggestions included adopting more exaggerated, cartoon-like gestures that would be clearer and more engaging for children. Communication style was another recurrent theme: long or repetitive instructions risked overloading some children, whereas others required concise prompts with additional support when necessary. The therapists converged on the need for adaptive communication strategies and for more expressive and socially oriented stimuli.

*C. Focus Group 3: Reflective Phase*

The final focus group, conducted at the end of the intervention cycle, encouraged a broader reflection on the overall trajectory of the project. Therapists acknowledged improvements in engagement and satisfaction, both for themselves and for the children, but stressed that the rigidity of scripted robot behaviors limited therapeutic flexibility. They argued that protocols should accommodate spontaneous contributions from children, such as personal narratives or unexpected gestures, and suggested mechanisms for branching storylines or corrective feedback where the robot could intentionally make mistakes. The group also underlined the importance of variability in reinforcement: repetitive praise was seen as quickly losing motivational value, while more nuanced social signals, including gaze and gesture, were regarded as crucial for sustaining authentic interaction.

*D. Cross-Group Observations*

Across the three sessions, one of the most notable patterns was the progressive increase in participation and openness among therapists. Over time, more individuals contributed actively, and feedback became increasingly specific and solution-oriented. This growth reflected not only familiarity with the robot but also trust in the process and confidence that their input would lead to tangible changes. The continuous design of the focus groups thus proved valuable in capturing evolving perspectives, deepening the dialogue between clinical and technical viewpoints, and consolidating a collaborative framework that extended beyond a single study.

## IV. DISCUSSION

The findings of this study indicate that continuous focus groups can address several limitations that have long been recognized in both HRI and clinical research. By engaging the same clinicians across multiple sessions, the method created space for reflections that evolved with practice, allowing participants to revisit earlier views and refine them in light of new experiences with the robot-assisted protocol. This dynamic stands in contrast to most prior uses of focus groups in HRI, which have typically been conducted as one-off events, producing valuable but static insights (Bradwell et al., 2021; Camilleri, Dogramadzi, & Caleb-Solly 2022). Continuity, in our case, fostered trust and a richer dialogue between researchers and clinical practitioners, thereby facilitating a more grounded integration of clinical expertise into the research process.

Calls for stronger methodological rigor in qualitative HRI resonate with these results. Concerns about inconsistent reporting and fragmented practices have been raised repeatedly (Veling & McGinn, 2021), echoing the principles of trustworthiness (Lincoln, Guba, & Pilotta, 1985) and elaborated through grounded theory approaches (Charmaz, 1995). By structuring data collection across several phases of an intervention, continuous focus groups embody criteria of credibility and transferability, while also operationalizing the reflexivity and transparency that Tracy (2010) identified as central to qualitative quality. Rigor in this sense does not depend on replication but on the ability to trace how knowledge is built iteratively and collaboratively.

The approach also extends discussions in participatory design. From early formulations that emphasized mutual learning (Bødker, 1994) to the later recognition of co-creation as a way of capturing tacit knowledge (Sanders & Stappers, 2008), participatory design has argued for more symmetrical relations between developers and stakeholders. Yet, as Spinuzzi (2025) noted, participation often falters when stakeholders are overburdened or lack the resources to remain engaged. This challenge is particularly acute in autism care, where families face systemic fragmentation and heavy caregiving demands (Gray, 2002; Mazurek et al., 2021; Bonfim et al., 2023). Continuous focus groups with clinicians offer a pragmatic solution: they do not replace families' voices but ensure consistent, professional engagement without adding new burdens to already strained households. In this sense, the method enacts a non-hierarchical, co-design stance, positioning therapists as collaborators who actively shape the trajectory of robot-assisted interventions.

Ethical considerations further underscore the value of this design. Research with vulnerable populations requires more than procedural compliance; it demands sensitivity to power relations, confidentiality, and autonomy. Guidance from the Association of Internet Researchers (AoIR) emphasizes that ethical decisions should be treated as contextual and iterative rather than fixed (Markham & Buchanan 2017). This view is reinforced by reflections on qualitative practice in health and education, which highlight the need for ongoing consent and reflexivity (Taquette & Souza 2022). Iteration in our focus groups played precisely this role: across sessions, participants had the opportunity to renegotiate their involvement, express new concerns, and revise earlier positions. Ethics thus became embedded in the process itself, not an external checkpoint, aligning with broader debates in HRI on responsible and socially aware research (Smakman, Vogt, & Konijn, 2021; Dautenhahn 2009).

The broader implications extend beyond autism. Interdisciplinarity has long been described as a defining feature of HRI (Goodrich & Schultz, 2007), yet, as Fraune et al. (2022) observe, collaboration across fields often remains additive rather than genuinely integrative. Continuous focus groups demonstrate how co-agency can be achieved in practice, by enabling therapists to bring their professional and tacit expertise into sustained dialogue with researchers and engineers. This approach also addresses the difficulty, noted in participatory design with older adults (Rogers, Kadylak, & Bayles, 2022), of maintaining stakeholder involvement over time. By creating a replicable model for longitudinal engagement, continuous focus groups provide a pathway for studying populations where direct participation is limited, such as children with ASD or older adults with dementia, while ensuring that technological design remains responsive to the realities of care and needs of the user.

Taken together, these findings advance both the practice of robot-assisted autism interventions and the methodology of HRI. A key implication of our findings concerns the necessity of mediators in interdisciplinary collaborations. While continuous focus groups created the structural conditions for dialogue, the actual translation of insights across clinical and technical vocabularies required professionals who could operate fluently in both domains. Without such figures, valuable clinical expertise risks being lost in translation, while technological possibilities may remain opaque to clinicians. We therefore argue that future HRI research in clinical contexts should explicitly cultivate and formalize interdisciplinary mediator roles—individuals who combine familiarity with therapeutic practice, sensitivity to ethical considerations, and technical literacy. These figures not only support iterative refinement of interventions but also safeguard against misalignments that can undermine the efficacy and acceptability of robot-assisted care.